\documentclass[prd,aps,showpacs,preprintnumbers,amssymb]{revtex4}

\usepackage{color}
\usepackage{epsf}

\begin{document}
\title{%
\hfill{\normalsize\vbox{%
\hbox{}
 }}\\
{Mass Ansatze for the standard model fermions  from a composite perspective}}

\author{Amir H. Fariborz
$^{\it \bf a}$~\footnote[1]{Email:
 fariboa@sunyit.edu}}

\author{Renata Jora
$^{\it \bf b}$~\footnote[2]{Email:
 rjora@theory.nipne.ro}}

 \author{Salah Nasri$^{\it \bf c}$~\footnote[3]{Email:
 nasri.salah@gmail.com}}

\affiliation{$^{\bf \it a}$ Department of Matemathics/Physics, SUNY Polytechnic Institute, Utica, NY 13502, USA}
\affiliation{$^{\bf \it b}$ National Institute of Physics and Nuclear Engineering PO Box MG-6, Bucharest-Magurele, Romania}
\affiliation{$^{\bf \it c}$ Department of Physics, College of Science, United Arab Emirates University, Al-Ain, UAE}

\date{\today}

\begin{abstract}
We consider a composite model in which the standard model fermions are bound states of  elementary spin $\frac{1}{2}$ particles, the preons,  situated in the conjugate product representation of the color group. In this framework we propose and analyze two mass Ansatze one for the leptons, the other one for the quarks, based on mass formulae of the Gell-Mann Okubo type. We find that these mass Ansatze can give an adequate  description of the known standard model fermion masses.
\end{abstract}
\maketitle

\section{Introduction}
The standard model of elementary particles has had remarkable success culminating with the discovery of the Higgs boson by the ATLAS \cite{Atlas} and CMS \cite{CMS} experiments. However there are still many theoretical unsolved issues like naturalness of the electroweak scale, the neutrinos masses and mixings, matter and antimatter asymmetry or the proliferation of fermions and fermion masses. It is this latter problem that we wish to explore in the present work. In general there are three types of answers to why there are so many generations of fermions in the standard model: a)  the number of fermions is accidental; b)  to consider some sort of gauge unification at a larger scale, enlarged group meaning larger representation of fermions that can accommodate all species; c)  the standard model fermions  are composite of more elementary fermions or bosons and fermions.  Here we shall reexamine the third explanation although arguably it lacks any experimental support and it is the most controversial.

Models of composite fermions have been proposed since the seventies in different frameworks and set-ups. One possibility exlored in the theoretical studies \cite{Pati}-\cite{Derman} is that the fermions are composite of a boson and a fermion. Another direction \cite{Mansouri}-\cite{Gatto} is to rely somewhat on gauge unfication and to build the fermions out of five constituents.  Finally another choice explored in the literature \cite{Terazawa}-\cite{Seiberg3}
is that the fermions are made out of only two constituents with or without the necessity of an enlarged gauge group.  All these studies have one feature in common: all the standard model fermions are built from a smaller number of constituents in one context or another.

The composite fermion models are plagued with all sorts of issues \cite{Peskin} including the absence of any known form factors that might be associated with the compositeness of leptons,  the possibility of exotic decay processes that have not been observed experimentally or the appearance in the effective theory of low dimension operators that might lead to rare processes unaccounted for.  We do not aim to resolve these issues in the present work
nor the problems that might appear with the dynamical assumptions but rather to consider a more practical approach: If the standard model fermions are to be composite at the electroweak scale then the composite nature of these objects should be able to account for or justify at least partially the proliferation of masses. Thus rather than the underlying dynamics and structure at higher scales a composite model should be able to give reasonable mass Ansatze inspired by what we know from another composite theory, that of baryons and mesons of low energy QCD.

In this work we construct the standard model fermions out of only two constituent preons situated in different combinations of the color group representations. We shall point out only briefly the dynamical assumptions at a larger scale and address in detail how the known masses of the standard model fermions may fit into a composite picture.

\section{Composite fermions}
We consider a composite model with two fermions $\Psi$ and $\chi$ that we should call generically preons (or psions and chions) situated in the conjugate product representation of the group $SU(3)_C$.  Since the product representation is reducible it will decompose into a direct sum of irreducible representations: $3^*\times3^*=3+6^*$. These preons have a global $U(1)_V\times U(1)_A\times SU(2)_L \times SU(3)_R$ symmetry with the assignment of the quantum numbers given in Table \ref{preons}.
\begin{table}[htbp]
\begin{center}
\begin{tabular}{c||c|c|c|c|c}
\hline \hline Preon   &$Q$ &$T_{3L}$  &$T_{3R}$&$SU(3)_C$
\\
\hline \hline $\Psi_L$ &$\frac{2}{3}$  &$\frac{1}{2}$   &$0$ &$3+6^*$
\\
\hline $\Psi_R$ &$\frac{2}{3}$  &$0$  &$\frac{1}{2}$ &$3+6^*$
\\
\hline $\chi_L$ &$-\frac{1}{3}$   &$-\frac{1}{2}$  &$0$ &$3+6^*$
\\
\hline $\chi_R$ &$-\frac{1}{3}$ &$0$   &$-\frac{1}{2}$ &$3+6^*$
\\
\hline \hline
\end{tabular}
\end{center}
\caption[Quantum number assignment for preons]{Quantum number assignment for preons } \label{preons}
\end{table}
Since we are dealing with Dirac fermions situated in  vector representations of the group $SU(3)_C$ this quantum number assignment is free of anomalies.

Assume that QCD with these fermion species and representations has the property of confining at a larger scale. Since the phase diagram for an $SU(N)$ gauge theory is mostly dictated by the behavior of the beta function this assumption is very plausible. Then it might be energetically favorable for two types of bound states to be formed as higher dimensional representations break into the lower ones: One type that are color singlets and another type that are color triplets. The actual phenomenon that might take place which we call "cascade confinement" will be discussed in a subsequent paper. Alternatively one may consider an additional very strong group $SU(3)_H$ (hypercolor) with preons in the adjoint representation that at a higher scale leads to the formation of hypercolor singlets.  However we shall not enter into detail here since our main purpose of the present work is to find reasonable mass Ansatze for fermion masses under the assumption that these are composite states of more elementary preons.

Besides scalar and fermion bound states vector and axial vector composite states may form. If the global symmetry is $SU(2)_L \times SU(2)_R$ then the two-preon bound states will be triplets and singlets under the global group by analogy with low energy QCD. We assume that at larger scale a parity violating operator and the formation of a vacuum condensate lead to the breaking of the group $U(1)_V\times U(1)_A \times SU(2)_L \times SU(2)_R$ down to $SU(2)_L\times U(1)_Y$. Since at this stage vector mesons with quantum number of triplets and singlets exist we can assume that this symmetry is gauged and the electroweak group is formed. The picture we just outlined may lack detail and may lead to dynamical issues but apart from these constitute a possible framework in which we will analyze the fermion masses.

The hypercharges associated with the preons after the spontaneous symmetry breaking of the large group down to the electroweak group are extracted simply from the relation:
\begin{eqnarray}
Y=2(Q-T_{3L})
\label{hyp665788}
\end{eqnarray}
and are summarized below:
\begin{eqnarray}
&&Y(\Psi_L)=\frac{1}{3}
\nonumber\\
&&Y(\Psi_R)=\frac{4}{3}
\nonumber\\
&&Y(\chi_L)=-\frac{1}{3}
\nonumber\\
&&Y(\chi_R)=-\frac{2}{3}.
\label{res5647}
\end{eqnarray}

Note that these are exactly the hypercharge assignment of the quarks in the standard model.

There are various singlet and triplet bound states that may form but here we will indicate only the scalar singlets that may be relevant to the spontaneous symmetry breaking of the electroweak group: one scalar doublet with hypercharge $Y=1$:
\begin{eqnarray}
\left(
\begin{array}{c}
\chi_R^{\dagger}\Psi_L\\
\chi_R^{\dagger}\chi_L
\end{array}
\right).
\label{one6645}
\end{eqnarray}
and the other with $Y=-1$:
\begin{eqnarray}
\left(
\begin{array}{c}
\Psi_R^{\dagger}\Psi_L\\
\Psi_R^{\dagger}\chi_L
\end{array}
\right).
\label{second554677}
\end{eqnarray}
These scalar doublets are doubled if one considers that for each of them there are two possible bindings $3\times 3^*$ or $6^*\times 6$.

We shall discuss separately the fermion bound states.

\subsection{Composite leptons}

The singlet three-preon bound states are given by the combinations $\bar{\Psi}\bar{\Psi}\bar{\chi}$ and $\bar{\chi}\bar{\chi}\bar{\Psi}$ first with charge $Q=-1$ that we associate with the charged leptons, second with the charge $Q=0$ that we associate with neutrinos.  By analogy with the baryons (neutron and proton) made of up and down quarks, namely by considering the antisymmetry of the wave function, it is impossible to form states of the type $\bar{\Psi}\bar{\Psi}\bar{\Psi}$ or $\bar{\chi}\bar{\chi}\bar{\chi}$ with spin $\frac{1}{2}$.

Next we need to determine the flavor structure of left handed and right handed states. We consider only one generation of leptons to find:
\begin{eqnarray}
&&e_L \,\,\, {\Psi}_R^{\dagger}{\Psi}_L^{\dagger}{\chi}_R^{\dagger}\,\,\, or\,\,\, {\Psi}_L^{\dagger}{\Psi}_L^{\dagger}{\chi}_L^{\dagger}
\nonumber\\
&&e_R\,\,\,{\Psi}_L^{\dagger}{\Psi}_R^{\dagger}{\chi}_L^{\dagger}\,\,\,or\,\,\,{\Psi}_R^{\dagger}{\Psi}_R^{\dagger}{\chi}_R^{\dagger}
\nonumber\\
&&\nu_L\,\,\, {\chi}_R^{\dagger}{\chi}_L^{\dagger}{\Psi}_R^{\dagger}\,\,\, or\,\,\, {\chi}_L^{\dagger}{\chi}_L^{\dagger}{\Psi}_L^{\dagger}
\nonumber\\
&&\nu_R\,\,\,{\chi}_L^{\dagger}{\chi}_R^{\dagger}{\Psi}_L^{\dagger}\,\,\,or\,\,\,{\chi}_R^{\dagger}{\chi}_R^{\dagger}{\Psi}_R^{\dagger}.
\label{assign5546747}
\end{eqnarray}
 First note that the right handed states and the left handed ones are related as it should by  application of  the parity operator. Second we will show that only the states in the right hand column in Eq. (\ref{assign5546747}) survive.  This stems from the fact that the total wave function must be completely antisymmetric. Since it is antisymmetric in color and symmetric in space it must be completely symmetric in flavor and spin. However a completely symmetric wave function in spin has the spin $\frac{3}{2}$  which means that the spin wave function must have mixed symmetry. Consequently the flavor wave function must have also mixed symmetry  and this requires various symmetrizations and antisymmetrizations in flavors. If for example one has the structure ${\Psi}_R^{\dagger}{\Psi}_L^{\dagger}{\chi}_L^{\dagger}$ where the last two preons form a scalar the symmetrization in flavor would require moving the last preon on various slots in the composite structure. But this would destroy the Lorentz properties of a composite state. Similar arguments indicate that no other states are possible. Thus the final structure for one generation of leptons is:
\begin{eqnarray}
&&e_L \,\,\, {\Psi}_L^{\dagger}{\Psi}_L^{\dagger}{\chi}_L^{\dagger}
\nonumber\\
&&e_R\,\,\,{\Psi}_R^{\dagger}{\Psi}_R^{\dagger}{\chi}_R^{\dagger}
\nonumber\\
&&\nu_L\,\,\,  {\chi}_L^{\dagger}{\chi}_L^{\dagger}{\Psi}_L^{\dagger}
\nonumber\\
&&\nu_R\,\,\,{\chi}_R^{\dagger}{\chi}_R^{\dagger}{\Psi}_R^{\dagger}.
\label{assign55467478}
\end{eqnarray}

Next we need to justify the three generations of leptons. This is done by  considering  various bindings that may form out of the representations of preons. Thus there are only three possibilities that lead to singlets:
\begin{eqnarray}
&& 3 \times 3\times 3=(6^s+3^*)\times 3=1+10+8
\nonumber\\
&&6^*\times 6^* \times 6^*=(15^*_s+15^*_a+6_s)\times 6^*=1+...
\nonumber\\
&&3\times 3\times 6^*=(6^s+3^*)\times 6^*_s=1+...,
\label{res664778}
\end{eqnarray} and each corresponds to a different choice of representations and thus to a different generation of leptons.

\subsection{Composite quarks}

We consider those fermionic colored bound states that can be formed in a lower representation of the color group (the triplet) and that are combinations of a scalar color singlet and triplet and a preon.  Note that this type of states are not uncommon in low energy QCD where it is presumed that the lower scalar and pseudoscalar mesons are mixings of two quark and four quark states, the latter being molecules of colorless two quark states. Thus we assume that the quarks are molecules of a colorless or colored scalar and a preon.

Generically the following states are possible: $\bar{\Psi}\bar{\Psi}\bar{\Psi}$, $\bar{\chi}\bar{\chi}\bar{\chi}$, $\bar{\Psi}\bar{\Psi}\bar{\chi}$ and $\bar{\chi}\bar{\chi}\bar{\Psi}$.  We will show that bound states that contain three preons of the same type cannot be formed. First we need to select the possible combinations of representations that lead to the desired bound states. The list narrows down to three possibilities:

1)
$ 3^*\times 3\times 3$
with two poossibilities:

a) $3^*\times 3\times 3=1\times 3+...=3+...  $ where the scalar state is formed by the first two preons and is a color singlet,

b) $3^*\times 3\times 3=3^*\times (3^*+6)=3+...$  where the scalar state is formed by the last two preons and is color triplet.

2) $6 \times 6^*\times3$
with two possibilities:

a) $6 \times 6^*\times 3=1\times 3+...=3+...$ where the scalar state is formed by the first two preons and is a color singlet

b) $6 \times 6^*\times 3=6\times (3^*+15^*)=3+...$  where the scalar state is formed by the last two preons and is color triplet.

3) Combined states:

a) $3^*\times 3\times 6^*=3^*\times(3^*+15^*)=3+...$  where the scalar state is formed by the last two preons and is color triplet.

b) $6\times 3\times 3=6\times (3^*+6)=3+...$ where the scalar state is formed by the last two preons and is color triplet.

These three possible grouping of representations correspond to three generations of fermions.

We shall discuss first the case where the scalar is formed by the two first preons in the combination. In this case the first two preons must form a spin singlet so the wave function is antisymmetric in spin.  Then the full spin wave function of the composite fermions has mixed symmetry in the second and third preons. Since the total wave function of the second and third preons must be totally antisymmetric and we cannot symmetrize in color(since this would destroy the singlet structure of the scalar formed out of the two first preons) we must symmetrize in flavor. This includes both symmetrization and antisymmetrization and cannot be done unless the preons on the second and third slots are of the same type (both left handed or right handed) and different flavors.

For the case when the scalar is formed out of the second and third preons this is an antisymmetric representation of color and is a spin singlet so it is both antisymmetric in spin and color. Since the wave function must be totally antisymmetric then it must be antisymmetric in flavor and again the second and third slots are of the same type (both left handed or right handed) and different flavors.

 Finally the following composite quark states emerge (where we depict only one generation of quarks the others being of the same type but with different bindings):

 \begin{eqnarray}
&& u_L\,\,\,\,{\rm is}\,\,\,\chi_L^{\dagger}\chi_R\Psi_R\,\,\,or\,\,\chi_L^{\dagger}\chi_L\Psi_L
 \nonumber\\
 &&u_R\,\,\,\,{\rm is}\,\,\,\chi_R^{\dagger}\chi_L\Psi_L\,\,\,or\,\,\, \chi_R^{\dagger}\chi_R\Psi_R
 \nonumber\\
 &&d_L\,\,\,{\rm is}\,\,\,\Psi_L^{\dagger}\chi_R\Psi_R\,\,\,or\,\,\Psi_L^{\dagger}\chi_L\Psi_L
 \nonumber\\
 &&d_R\,\,\,\,{\rm is}\,\,\,\Psi_R^{\dagger}\chi_L\Psi_L\,\,\,or\,\,\, \Psi_R^{\dagger}\chi_R\Psi_R.
 \label{reserte64785}
 \end{eqnarray}
Note that the combinations on the third column in the equation above can exist only when the scalar in the composite quark is formed out of the second and third preons.

\section{Mass Ansatze for charged leptons and neutrinos}

Charged leptons and neutrinos have the same structure as the baryons so we explore whether formulas similar to the Gell Mann Okubo formula may apply  to them. Therefore we assume that the masses of the charged leptons are given by:

\begin{eqnarray}
&&m_e=b_1+2m_1+m_2
\nonumber\\
&&m_{\mu}=b_2+2m_1+m_2
\nonumber\\
&&m_{\tau}=b_3+2m_1+m_2
\label{res546775}
\end{eqnarray}
where $m_1=m_{\Psi}$, $m_2=m_{\chi}$ and $b_1$, $b_2$, $b_3$ correspond to the three binding energies corresponding to the three combinations of representations. Here the intrinsic assumption is made that after spontaneous symmetry breaking the preons will get constituent masses no matter how large.

Similarly the Dirac masses of the three neutrinos are given by:
\begin{eqnarray}
&&m_{\nu_e}=b_1+2m_2+m_1
\nonumber\\
&&m_{\nu_{\mu}}=b_2+2m_2+m_1
\nonumber\\
&&m_{\nu_{\tau}}=b_3+2m_2+m_1.
\label{masret5464}
\end{eqnarray}

However neutrinos have no charge and they may have Majorana masses. We assume that at a higher scale where the symmetry $U(1)_V\times U(1)_A \times SU(2)_L \times SU(2)_R$ is broken and the parity violating operator acts large right handed neutrino Majorana masses are generated (note that the right handed neutrinos contain only right handed preon states) corresponding to the breaking of the $SU(2)_R$ symmetry.

Then the neutrino masse are obtained in a standard way through a see-saw mechanism which we shall review briefly here. The full mass term Lagrangian has the form:
\begin{eqnarray}
{\cal L}_m=D_{ij}\bar{\nu}_i\nu_j+B_{ij}\bar{\nu}^c_{Ri}\nu_{Rj}+h.c.
\label{lagr54664}
\end{eqnarray}

We reexpress the neutrino states $\bar{\nu}_i$ and $\nu_i$ in terms of the Majorana states $q_i$, $\omega_i$.:
\begin{eqnarray}
&&q_i=\nu_{Li}+\nu_{Li}^c
\nonumber\\
&&\omega_i=\nu_{Ri}+\nu_{Ri}^c.
\label{res5463534}
\end{eqnarray}
In terms of the latter states the Lagrangian will become:
\begin{eqnarray}
{\cal L}_m=\frac{1}{2}D_{ij}(\bar{q}_i\omega_j+\bar{\omega}_iq_j)+B_{ij}\bar{\omega}_i\omega_j
\label{newre546}
\end{eqnarray}

Let us assume for simplicity that neutrinos are degenerate ($D_{ij}=D\delta_{ij}$, $B_{ij}=B\delta_{ij}$) so they have the same masses. Then there are two mass eigenstates in first order $B\gg D$):
\begin{eqnarray}
&&m_{\nu1}=-\frac{D^2}{4B}
\nonumber\\
&&m_{\nu 2}=B+\frac{D^2}{B}.
\label{mass435553}
\end{eqnarray}
Then it is clear that one very small set of masses ($m_{\nu1}$) emerges and one very large $m_{\nu 2}$. The mixing angle is $tan(\theta)\approx \frac{-D}{B}\approx 0$ and the eigenvalues will be $q_i$ corresponding to the small mass and $\omega_i$ corresponding to the large one. Then the Majorana fermions with small masses are mostly left handed states that couple with the electroweak sector whereas the Majorana fermions with very large masses are mostly right handed states that decouple from the electroweak sector.

\section{Mass Ansatz for the quarks}

For the quark masses we shall consider a combination of the Gell Mann Okubo formula applied to scalars and composite fermions. We need to take into account the four possible bindings ( or contribution to the scalar masses) that we denote as follows: $a_1^{1/2}$ for the binding $3^* \times 3$ that leads to the formation of the singlet; $a_2^{1/2}$ the binding energy associated to $6^*\times 6$ that again leads to a scalar singlet; $b_1$  for the binding $3^* \times 6$ that leads to the formation of a scalar triplet; $b_2$ for the binding $3\times 3$ that leads to the formation of a scalar triplet. Since the last two cases also combine the scalar with the third preon we will use $b_1$ and $b_2$ also for this purpose.  In the end the following quark mass Ansatz emerges:

\begin{eqnarray}
&&m_u=\frac{1}{3}\Big[\sqrt{a_1+2m_2s}+\sqrt{a_1+(m_1+m_2)s}+m_1+2m_2+\sqrt{b_2^2+(m_1+m_2)s}+b_2\Big]
\nonumber\\
&&m_d=\frac{1}{3}\Big[\sqrt{a_1+2m_1s}+\sqrt{a_1+(m_1+m_2)s}+m_2+2m_1+\sqrt{b_2^2+(m_1+m_2)s}+b_2\Big]
\nonumber\\
&&m_c=\frac{1}{3}\Big[\sqrt{a_2+2m_2s}+\sqrt{a_2+(m_1+m_2)s}+m_1+2m_2+\sqrt{b_1^2+(m_1+m_2)s}+b_1\Big]
\nonumber\\
&&m_s=\frac{1}{3}\Big[\sqrt{a_2+2m_1s}+\sqrt{a_2+(m_1+m_2)s}+m_2+2m_1+\sqrt{b_1^2+(m_1+m_2)s}+b_1\Big]
\nonumber\\
&&m_t=\frac{1}{2}\Big[\sqrt{b_1^2+(m_1+m_2)s}+\sqrt{b_2^2+(m_1+m_2)s}+b_1+b_2+2m_2\Big]
\nonumber\\
&&m_b=\frac{1}{2}\Big[\sqrt{b_1^2+(m_1+m_2)s}+\sqrt{b_2^2+(m_1+m_2)s}+b_1+b_2+2m_1\Big].
\label{mass435534}
\end{eqnarray}
Here $s$ is a scale that establishes the correct mass dimensionality and $m_1=m_{\Psi}$, $m_2=m_{\chi}$. Note that all parameters including the masses may be positive or negative.

The  equations in (\ref{mass435534}) further lead to the following relations:

\begin{eqnarray}
&&3(m_u-m_d)=\sqrt{a_1+2m_2s}-\sqrt{a_1+2m_1s}+m_2-m_1
\nonumber\\
&&3(m_c-m_s)=\sqrt{a_2+2m_2s}-\sqrt{a_2+2m_1s}+m_2-m_1
\nonumber\\
&&m_t-m_b=m_2-m_1
\nonumber\\
&&m_t=
\nonumber\\
&&\frac{1}{2}\Big[3(m_u+m_c)-2m_1-2m_2-\sqrt{a_1+2m_2s}-\sqrt{a_1+(m_1+m_2)s}-\sqrt{a_2+2m_2s}-\sqrt{a_2+(m1+m_2)s}\Big].
\label{res443564778}
\end{eqnarray}

We denote:
\begin{eqnarray}
&&A=3m_c-\sqrt{a_2+2m_2s}-\sqrt{a_2+(m_1+m_2)s}-2m_2-m_1
\nonumber\\
&&B=3m_u-\sqrt{a_1+2m_2s}-\sqrt{a_1+(m_1+m_2)s}-2m_2-m_1
\label{res433242}
\end{eqnarray}
to find that:
\begin{eqnarray}
&&b_2=\frac{B^2-(m_1+m_2)s}{2B}
\nonumber\\
&&b_1=\frac{A^2-(m_1+m_2)s}{2A}.
\label{res332324}
\end{eqnarray}

\begin{figure}
\begin{center}
\epsfxsize = 10cm
\epsfbox{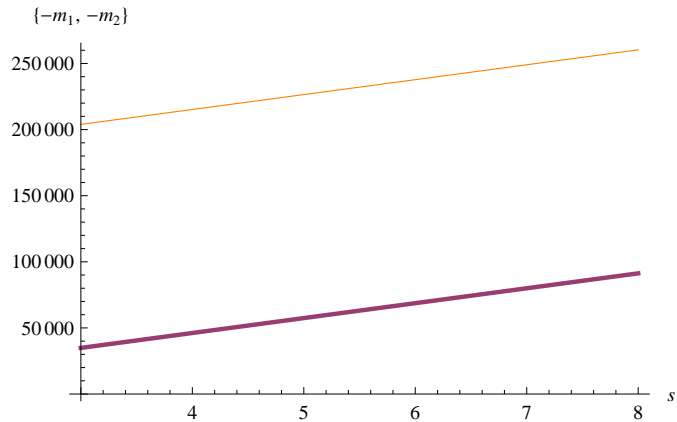}
\end{center}
\caption[]{%
Plot of the negative masses of the constituent preons as a function of the scale $s$; here $m_1=m_{\Psi}$ (orange line) , $m_2=m_{\chi}$ (thick line). The masses  and the parameter $s$ are expressed in MeV.
}
\label{masses}
\end{figure}

\begin{figure}
\begin{center}
\epsfxsize = 10cm
\epsfbox{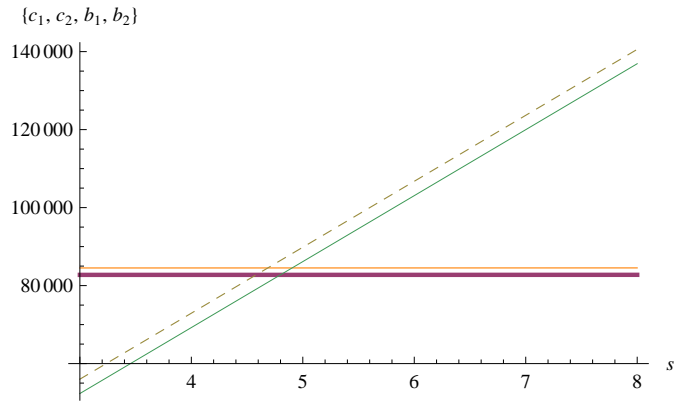}
\end{center}
\caption[]{%
Plot of the binding energies $c_1$ (orange line), $c_2$ (thick line), $b_1$ (dashed line), $b_2$ (thin line) as a function of the scale $s$. The binding energies and $s$ are expressed in MeV.
}
\label{bindings}
\end{figure}

From Eqs. (\ref{res443564778}) and (\ref{res332324}) we were able to determine the parameters $c_1=a_1^{1/2}$, $c_2=a_2^{1/2}$, $b_1$, $b_2$, $m_1$ and $m_2$ as a function of the scale $s$. The results are plotted in the Figs \ref{masses} and \ref{bindings}.

In Eq. (\ref{mass435534}) we have $6$ equations and $7$ parameters so it seems that our mass Ansatz is somehow arbitrary and unfit to make any prediction. However we expect that the binding energies should be very close in value so for our Ansatz to be meaningful there should be a small region of the parameter $s$ where all these binding energies have the same approximate values. That this is indeed the case for $ 4.5 \,{\rm MeV}<s<5\,{\rm MeV}$  as it can be read off clearly from Fig. \ref{bindings}. This means that in first order approximation the six quarks masses are determined from essentially four parameters; the masses of the constituent preons, the common binding energy and the scale.

\section{Discussion and conclusions}

In this work we considered the possibility that the standard model fermions may be bound states of more three constituents, the preons. In total there are only two preons situated in the conjugate product representaions of the color group. What distinguishes the three generation of fermions is the combination of group reprsentations that participate to the fermion structure. The model is simple and economical and has some predictive power.

Rather than dwelling on the dynamical issues related to the higher scale in the theory we focused on the composite fermion masses derived from their underlying structure and inspired by what we know from low energy QCD. We thus apply Gell-Mann Okubo type mass relations to both the leptons and quarks by analogy to similar mass relations that exist for baryons and pseudoscalar mesons. We find that the composite picture makes sense for both leptons and quarks.  The mass differences between different generations of leptons are due to different binding energies whereas those between charged leptons and neutrinos are due to a different preon structure.  The smallness of the neutrino masses is associated with the existence of a  large Majorana mass for the right handed neutrinos  at the scale where the original group is broken and thus is derived through  a see-saw mechanism.

The quark masses are more intricate as in this case the Gell-Mann Okubo formula is applied both to the constituent scalar that forms in their structure and to the whole composite fermion. By plotting the binding energies against the scale that appears in the mass formulae one observes that for a scale in the range $ 4.5 \,{\rm MeV}<s<5\,{\rm MeV}$ the binding energies are similar in values showing that in the first order approximation there is a picture where all binding energies are the same  and the six quark masses can be described only in terms of four parameters;  the masses of the constituent preons, the common binding energy and  the scale.

It is assumed that before spontaneous symmetry breaking the dynamics and group structure is such that all composite fermions have the same binding energy and constituent preon masses such the overall standard model fermion masses are zero. This can be ensured by a 't Hooft type of mechanism of manifest chirality conservation.
However the purpose of our paper was not to analyze various dynamical issues associated to composite theories in general but rather to examine and analyze meaningful mass Ansatze that make sense in a particular composite model.

\section*{Acknowledgments} \vskip -.5cm

The work of R. J. was supported by a grant of the Ministry of National Education, CNCS-UEFISCDI, project number PN-II-ID-PCE-2012-4-0078.

\end{document}